\documentclass[a4paper,12pt]{article}

\def\ad   {a^{\dagger}}

\def\bn   {\overline {n}}

\def\HP   {\hat {P}}

\def\al   {\alpha}
\def\be   {\beta}

\def\Gam  {\Gamma}

\def\del  {\delta}
\def\Del  {\Delta}

\def\HA {\hat A}

\def\HH {\hat H}

\def\HJ {\hat J}
\def\hj {\hat j}
\def\HK {\hat K}

\def\HP {\hat P}

\def\HU {\hat U}

\def\HH {\hat H}
\def\HK {\hat K}

\def\Heta {\hat \eta}
\def\oc {\overline c}

{\it}
{\it}
{\it}
{\it}
{\it}
{\it}
\def\be {\beta}

\def\de {\delta}

\def\sig {\sigma}

\def\om {\omega}

\usepackage{times}
\usepackage{graphicx}
\begin{document}
\title{  Effect of residual many-body forces due to the evolution
in the in-medium similarity renormalization group method.}
\author{G. Puddu\\
       Dipartimento di Fisica dell'Universita' di Milano,\\
       Via Celoria 16, I-20133 Milano, Italy}
\maketitle
\begin {abstract}
      In the past few years  in-medium similarity renormalization group methods
      have been introduced and developed. In these methods the Hamiltonian is evolved
      using a unitary transformation in order to decouple a reference state from the
      rest of the Hilbert space. The evolution by itself will generate, even if we start from a
      two-body interaction, many-body forces which are usually neglected. In this work
      we estimate the effect of these residual many-body forces by comparing results obtained
      with  the Hybrid Multi-determinant
      method, which keeps the Hamiltonian within the two-body sector, with the corresponding ones
      obtained with the in-medium similarity renormalization group. 
      Although percentage-wise the effect of neglecting these induced many-body forces
      is not too large, they can be appreciable depending on the nucleus, the shell model space
      and the harmonic oscillator frequency.

\par\noindent
{\bf{Pacs numbers}}: 21.60.-n,21.60.De
\par\noindent
\vfill
\eject
\end{abstract}
\section{ Introduction.}
\bigskip
\par
      In the past years we have witnessed the development of powerful ab-initio many-body
      techniques to solve the nuclear Schroedinger equation. Among these methods we mention
      the no-core shell model (NCSM) (refs. [1]-[4]), the coupled-cluster method (refs.[5]-[8])
      and the in-medium similarity
      renormalization group (IM-SRG) (see refs.[9],[10] for comprehensive reviews and references in there).
      As well known  the limitation of the NCSM is 
      the size of the Hilbert space. However, where applicable, the NCSM  gives exact results.
      In the IM-SRG approach the many-body Hamiltonian is transformed with a unitary many-body
      operator. The discussion that follows is applied only to two-body forces and
      we shall consider only closed shell nuclei. 
\par
      The key idea of the IM-SRG is as follows.
      The original Hamiltonian is first rewritten in normal form with respect to a reference state.
      In order to achieve this the formalism of Kutzelnigg and Mukherjee (ref.[11]) is used by which the
      original Hamiltonian, written with the particle vacuum as a reference state, is rewritten
      with a different reference state as a vacuum. For closed shell nuclei, the simplest choice is to use
      a single Slater determinant as a reference state, usually the spherical Hartree-Fock solution.
      The Hamiltonian acquires a zero-body term, a normally ordered one-body term and 
      the normally ordered two-body interaction.
      The next step is the definition  of a unitary operator which transforms this
      new Hamiltonian into a  Hamiltonian such that 
      the reference state is decoupled from the rest of the Hilbert space.
      This is achieved via a flow equation. There are several choices for this unitary operator
      (cf. refs.[9],[10]).
      This parameter-dependent flow equation will give, at the end, the ground-state energy
      as the the zero-body coefficient. Effectively the flow will transform the original
      Hamiltonian into a block diagonal form with the ground-state decoupled from the rest of the Hilbert
      space.
      As well known, the flow will also generate induced many-body interactions
      which are normally neglected (this is the so-called IM-SRG(2) truncation).
      Hence, although we started with  a two-body interaction only, the exact flow is approximated
      at the IM-SRG(2) level discarding all induced many-body interactions. 
      It is natural to address the question about the accuracy of this truncation.
      Ideally we would like to compare IM-SRG(2) (which will call simply IM-SRG from now on)
       results with methods that do not generate
      induced many-body forces.
      An ideal  method would be to compare the IM-SRG results with 
      the shell model diagonalization in the full Hilbert space
      using Lanczos methods. However this can be done for light nuclei and not too large
      single-particle spaces. Here we use the Hybrid Multi-Determinant method (HMD)(refs.[12],[13])
      complemented
      by energy versus variance of energy extrapolation techniques (EVE) (refs.[14]-[20]).
       Although computationally
      more demanding than the IM-SRG, it does not generate induced many-body forces, and
      in principle its applicability does not depend on the size of the Hilbert space.
      The key idea of the HMD method is to expand the nuclear wave function as a linear
      combination of many generic Slater determinants (with exact or approximate restoration
      of good quantum numbers using projectors) and to determine these Slater
      determinants using energy
      minimization techniques. The final EVE step consists in the evaluation of
      the energy variance (or related quantities) $ < ( \HH -<\HH>) ^2 >$, $\HH$ being
      the two-body Hamiltonian, once a set of approximate wave functions has been
      determined. The energy has a linear or linear+quadratic behavior as a function
      of the above variance. Extrapolation to zero variance will give the ground state energy.
      Although applicable, the HMD method becomes more and more computationally intensive
      as we increase the single-particle space. Therefore we consider, in this work,
      Hamiltonians generated from the bare Hamiltonian for the nucleus of interest by the
      Lee-Suzuki (LS) renormalization technique (refs.[21]-[24]).
      Alternatively one could evolve first
      the two-body interaction with the similarity renormalization group in the vacuum,
      in order to soften the two-body interaction (refs.[25]-[28]). This preliminary step defines the 
      Hamiltonian in the two-body sector which we use to perform the comparison. 
\par
      The outline of this paper is as follows. In section 2 we briefly discuss the Hamiltonian
      we consider. In section 3 we recall briefly the HMD method, in section 4 we outline the
      spherical IM-SRG flow equations (see also ref.[10]) using the
      Brillouin generator (ref.[9]). In section 5 we compare the results in an harmonic oscillator
      basis  for ${}^4He$ and for ${}^{16}O$, and in section 6 we give some conclusions. 
\bigskip
\section{ The Hamiltonian.}
\bigskip
\par
      We start from the two-body Hamiltonian
$$
\HH= \sum_{i=1}^A {p_i^2\over 2m}+\sum_{i<j} V_{ij}
\eqno(1)
$$
      where $m$ is the nucleon mass and $V_{ij}$ is the interaction between 
      nucleons $i$ and $j$. Here we consider the chiral $N3LO$ interaction of Entem and Machleidt (ref.[29]).
      Much in the same way it is done in the NCSM (ref. [2]) we add
      a confining harmonic oscillator potential for the center of mass and obtain
      an A-dependent Hamiltonian. If $\om$ is the frequency of this potential, the
      Hamiltonian can be recast in the following form
$$
\HH_{\om}=\HH+{1\over 2} mA\om^2 R_{c.m.}^2=\sum_{i=1}^A h_i +\sum_{i<j}  V_{ij}^{(A)}
\eqno(2)
$$
     with
$$
 V_{ij}^{(A)}=V_{ij} -{m \om^2\over 2 A}(\vec r_i-\vec r_j)^2
\eqno(3)
$$
      and
$$
h_i= {p_i^2\over 2 m} + {1\over 2} m \om^2 r_i^2
\eqno(4)
$$
     At the level of the 2-cluster approximation (ref.[2]), the 2-particle Hamiltonian 
     which is the input of the LS method, is the intrinsic part of 
$$
\HH_{12} =  h_1+h_2 +V_{12}^{(A)}= h_{rel}+ h_{cm}
\eqno(5)
$$
     The intrinsic A-dependent two-particle Hamiltonian  $h_{rel}$ in eq.(5) 
     is renormalized.
     First all matrix elements of $h_{rel}$ are evaluated in the center of mass frame
     using no less than   $200$ major shells. The radial integrals were evaluated using
     $3000$ integration points.
     We then renormalize this Hamiltonian 
     using, as the $P$ space, all relative-momentum HO states having pair quantum numbers
     $ n,l$ satisfying $2n+l\leq 2 N_{lab}$. The $Q$ space is comprised of all HO states not included
     in the $P$ space.
     For numerical stability we use the method due to Kvaal (ref.[30]).
\par 
     We augment the degrees of freedom by including the center of mass degrees of freedom and
      then translate this Hamiltonian to the lab frame using the Talmi-Moshinski brackets (cf. 
     ref. [31] for a numerically efficient method and available subroutines).
     The single-particle space in the lab frame is limited to
     $2 n+l\leq N_{lab}$. We consider here only an harmonic oscillator basis.
     For the  details of the implementation see ref. [32] (especially the appendix).
      In principle, if carried out exactly, the renormalization prescription
     should generate multiparticle effective potentials. We discard all induced many-body interactions
     and define our model Hamiltonian solely in the two-body sector. We do not pretend that this Hamiltonian
     to accurately describe the nucleus under consideration, we simply stress that we use the above prescription as
     a definition of the two-body Hamiltonian for the nucleus.
     This Hamiltonian may or may not include
     the harmonic oscillator center of mass term $\be(\HH_{cm}-3\hbar\om/2)$.
     In order to simplify as much as possible the comparison with the IM-SRG method we consider
     $N_{lab}$ as low as possible. Although large $N_{lab}$ values can be handled by the IM-SRG method,
     it would be problematic to compare IM-SRG results with those which work strictly within
     the two-body sector, since we wish to  assess the importance of the many-body forces omitted in
     the IM-SRG flow.
\bigskip
\section{ A brief recap of the HMD method.}
\bigskip
\par
     The Hybrid-Multi-Determinant (HMD) method consists in generating more and more accurate
     wave-functions with  linear combinations of many
     particle-number conserving Slater determinants 
     which  are determined with quasi-Newtonian energy minimization techniques
     (cf. ref. [33]). We use a rank-3 update technique described in details in ref.[34].
     The wave-function of the nucleus is written as
$$
| \psi>= \sum_{S=1}^{N_D} g_S \HP |U_S>
\eqno(6)
$$
       where $\HP$ is a projector to good quantum numbers (e.g. good angular momentum and parity)
       $N_D$ is the number of Slater determinants $|U_S>$ expressed as
$$
|U_S> = \oc_1(S)\oc_2(S)... \oc_A(S) |0>.
\eqno(7)
$$
       The generalized creation operators $\oc_{\alpha}(S)$ for $\al=1,2,..,A$ are a linear combination
       of the creation operators $\ad_i$
$$
\oc_{\al}(S)=\sum_{i=1}^{N_s}U_{i,\al}(S)\ad_i  \;\;\;\;\;\al=1,...A
\eqno(8)
$$
       Here $N_s$ is the number of the single-particle states.
       The complex coefficients $U_{i,\al}(S)$ represent the single-particle wave-function of the
       particle $\al=1,2,..,A$. We do not impose any symmetry on the Slater determinants (axial or other)
       since the $U_{i,\al}$  are variational parameters. 
       These complex coefficients are obtained by minimizing the energy expectation values
$$
E[U]= { <\psi |\HH |\psi> \over <\psi |\psi>}
\eqno(9)
$$
       The coefficients $g_S$ are obtained by solving the generalized eigenvalue problem
$$
\sum_{S} <U_{S'} |\HP\HH | U_S> g_S = E \sum_{S} <U_{S'}|\HP| U_S> g_S  
\eqno(10)
$$
       for the lowest eigenvalue $E$. We normally consider projectors to good $z-$ component of
       the angular momentum and parity, rather than projectors to good angular momentum and
       parity.
       The number of Slater determinants necessary for convergence can be quite large. Hence
       as a final step we use EVE techniques (we implement the variant of ref.[19]).
       These techniques consist  in the evaluation of the energy variance 
$$
\sig^2=<\HH^2>-<\HH>^2
\eqno(11)
$$
       as a function of the energy $<\HH>$. On general grounds $<\HH>$ is a linear or
       linear+quadratic function of $\sig^2$ provided we are sufficiently close to the ground-state
       (here we are primarily concerned about ground-state energies). The clause "sufficiently close"
       is basic. In fact, we extrapolate the plot $<\HH>$ as a function of $\sig^2$ to $\sig^2=0$.
       Even small uncertainties in the coefficients of the expansion can cause sizable error
       in the extrapolated ground-state energies if we are not sufficiently close to 
       $\sig^2=0$. Differently stated, the linear+quadratic fit must be of
       high quality. In practice, we generated a few hundreds  Slater determinants as sketched
       above,  we then evaluate $\sig^2$ and $<\HH>$ for a partial linear combination until
       we include them all. As a final step we employ the linearization method introduced
       in ref. [18] to optimize the order of the Slater determinants,
        More precisely, we construct a set of point $(\sig^2,<\HH>)$
       using wave-functions $| \psi>= \sum_{S=1}^N g_S \HP |U_S>$ for all $N\leq N_D$
       after these Slater determinants have been reordered so that the points $(\sig^2,<\HH>)$
       can be fitted with a linear+quadratic curve. 
\bigskip
\section{ A brief description of the IM-SRG method.}
\bigskip
       Self-contained and detailed descriptions of the IM-SRG method and its applications
       can be found in the review papers of refs. [9],[10]. 
       Here we simply describe how it has been implemented in this work.
       The basic idea is to evolve the many-body Hamiltonian with a continuous set 
       of unitary operators $\HU(s)$
$$
\HH(s) = \HU(s) \HH \HU(s)^{\dagger}
\eqno(12)
$$
       The flow equation is then given by
$$
{d\HH(s)\over ds} = [\Heta(s),\HH(s)]
\eqno(13)
$$
       where the generator $\Heta(s)$ is given by $\Heta(s)={d\HU(s)\over ds}\HU(s)^{\dagger}$.
       An approximate ground-state wave-function $|\phi>$ is selected and the initial Hamiltonian
       is rewritten in normal form with respect to $|\phi>$ as a vacuum. Both the generator and
       the evolved Hamiltonian are truncated at the two-body level as
$$
\Heta(s) = \sum_{ij}\eta^i_j(s) :\HA^i_j: + (1/4)\sum_{ijkl}\eta^{ij}_{kl}(s):\HA^{ij}_{kl}:
\eqno(14)
$$
$$
\HH(s)= E(s)+\sum_{ij}f^i_j(s) :\HA^i_j:+ (1/4)\sum_{ijkl}\Gam^{ij}_{kl}(s):\HA^{ij}_{kl}:
\eqno(15)
$$
       Here we use the tensor notation commonly employed in the IM-SRG method,
        that is $\HA^i_j=\ad_i a_j$ and
       $ \HA^{ij}_{kl}=\ad_i\ad_j a_l a_k$. As usual, the colons denote normal ordering 
       to the reference state $|\phi>$.
       We use the Brillouin generator since ${dE(s)\over ds}$ assumes a particularly appealing form.
       Since we study only closed shell systems we use a single reference $|\phi>$, as usually done
       for closed (sub)shells. In order to save computer memory we work in the angular momentum 
       coupled representation. We solve the flow equations ${dE(s)\over ds},\;\;{df(s)\over ds}$
       and ${d\Gam(s)\over ds}$
       using the Runge-Kutta method of rank 3 (cf. ref.[35])) with $\Del s=0.000025$
       until the energy no longer changes. The relevant flow equations in the coupled 
       representations  are described in the Appendix, using as a basis the natural orbit
       representation that diagonalized the expectation values of the one-body density.
       Three-body and higher rank terms arise from the commutator in eq.(13) and they are
       unavoidable in this method. In principle these depend of the choice of the reference
       state. As discussed in the next section, we in particular want to study the eventual discrepancy
       between the HMD method in which the whole Hamiltonian remains in the two-body sector and
       the IM-SRG results. We expect a dependence on the reference state, on the harmonic oscillator
       frequency $\hbar\om$ and on the size of the single particle space defined by $N_{lab}$. We will use
       two single-reference states. The spherical Hartree-Fock solution and a naive filling of the lower
       harmonic oscillator orbits. 
       Note that in section 5 of ref.[10] 
       for ${}^4He$ a discrepancy between IM-SRG binding energy and the one obtained with
       the Fadeev-Jakubowski method has already been found a bit large, pointing out to a sizable
       effect of the neglected induced many-body forces in the evolution.
       Presumably these discrepancies can be reduced using the more involved 
       multi-reference states as done recently in ref.[36].
\bigskip
\section{ Numerical results.}
\bigskip
\par
       We considered the cases of ${}^4He$ and ${}^{16}O$.
       For  ${}^4He$ we considered the following cases. The harmonic oscillator frequencies (in MeV's) 
       are $\hbar\om=24,\;\;38$ for $N_{lab}=3,4,5$. For ${}^{16}O$,
       $\hbar\om=14\;$ with  $N_{lab}=3,4,5$, for $\hbar\om=24\; N_{lab}=2,3,4$ and 
       for $\hbar\om=32$ with $N_{lab}=2,3$.
       It should be pointed out that changing $N_{lab}$ the Hamiltonian changes. It does not
       correspond to a different truncation. Hence we do not expect a monotonic behavior of the
       energies as we increase $N_{lab}$.
       The largest size of the Hilbert space for $J_z^{\pi}=0^+$ is about $10^{22}$
       for ${}^{16}O$ with $N_{lab}=5$, while for $N_{lab}=3$ and $N_{lab}=4$ the size of
       the Hilbert space is $4\times 10^{14}$ and
       $5\times 10^{18}$ respectively. 
       For ${}^4He$ with $N_{lab}=5$ the corresponding number is $3.529.304$.
       In the ${}^{16}O$ cases we added to the Hamiltonian a center of mass term with $\be=1 MeV$. 
       The results are summarized in the tables.
       Let us discuss first the ${}^4He$.
       The HMD results for ${}^4He$ are rather accurate. An estimate of the uncertainty for $N_{lab}=3,4,5$
       is about a  dozen  $KeV$'s. We considered two reference states for the IM-SRG calculations.
       One is the naive Fermi filling (FF) of the lowest single-particle states and 
       the other is the spherical HF.
       Notice that for the lowest value of $\hbar\om=24MeV$ the discrepancies between the various methods
       increases with $N_{lab}$. This discrepancy increases for the largest value of $\hbar\om$.
       This discrepancy is substantial and is qualitatively in agreement with the findings of ref. [10]
       where the IM-SRG results for large single-particle spaces have been compared with the exact
       binding energy for this interaction. 
\renewcommand{\baselinestretch}{1}
\begin{table}
   \begin{tabular}{| c | c | c| c| c | }
          \hline
 $\hbar\om(MeV)$  & $ N_{lab}$ & HMD & FF(MeV) & HF(MeV)  \\
          \hline
  24   &     3      &   -25.798    &  -25.933   &  -25.932 \\
       &     4      &   -24.670    &  -24.676   &  -24.975 \\
       &     5      &   -24.127    &  -24.261   &  -24.888 \\
          \hline
  38   &     3      &   -22.047   &   -21.145   &  -22.539 \\
       &     4      &   -23.027   &   -21.989   &  -24.268 \\
       &     5      &   -23.544   &   -22.264   &  -24.907 \\
          \hline
\end{tabular}
\caption { Ground-state energies for ${}^4He$ obtained with the HMD method,
 with the IM-SRG with the Fermi filling (FF) and HF as a reference state.
 The experimental value is $-28.295$ MeV (ref.[37])}
\end{table}
\renewcommand{\baselinestretch}{1}
\begin{table}
   \begin{tabular}{| c | c | c| c| c | }
          \hline
 $\hbar\om(MeV)$  & $ N_{lab}$ & HMD & FF(MeV) & HF(MeV) \\
          \hline
  14   &     3      &  -149.634   &  -153.646   &  -148.680 \\
       &     4      &  -139.826   &  -142.858   &  -138.301 \\
       &     5      &  -133.397   &  -135.467   &  -130.081\\
          \hline
  24   &     2      &  -139.706   &   -139.699  &  -139.704 \\
       &     3      &  -113.057   &   -110.416  &  -111.424 \\
       &     4      &   -97.855   &   -93.243   &  -95.664  \\
          \hline
  32   &     2      &   -53.921   &   -54.138   &  -54.185 \\
       &     3      &   -62.823   &   -58.461   &  -61.097 \\
          \hline
\end{tabular}
\caption { Ground-state energies for ${}^{16}O$ obtained with the HMD method,
 with the IM-SRG with the Fermi filling (FF) and HF as a reference state.
The experimental value is $-127.619$MeV (ref.[37])}
\end{table}
      It is instructive to plot the results for $E(s)$ at large $s$ obtained with the IM-SRG
      with the corresponding ones obtained with the HMD. In figs. (1)-(6) we show this comparison.
      For the ${}^4He$ cases, the uncertainties of the HMD method are best quantified by
      plotting $E(\sig^2)$ vs. $\sig^2$ together with the linear+quadratic fit.
      This is done in figs. (7)-(8) for $N_{lab}=5$.
      Notice that we considered only the end part (i.e. the one
      closer to the vertical axis) of the data points.
      The data points are close to the energy axis, and in such  cases
      the extrapolated values are accurate.
      The case of ${}^{16}O$ is less certain. In principle we can get close to the energy axis in the
      EVE plot, but we would have to consider a large number of Slater determinants.
      In these cases we can estimate the uncertainty in the HMD method by performing 
      several fits to different sets of $(\sig^2,E(\sig^2))$ data points. For some of the best fits
      the extrapolated values are shown in table 2.
\renewcommand{\baselinestretch}{1}
\begin{figure}
\centering
\includegraphics[width=10.0cm,height=10.0cm,angle=0]{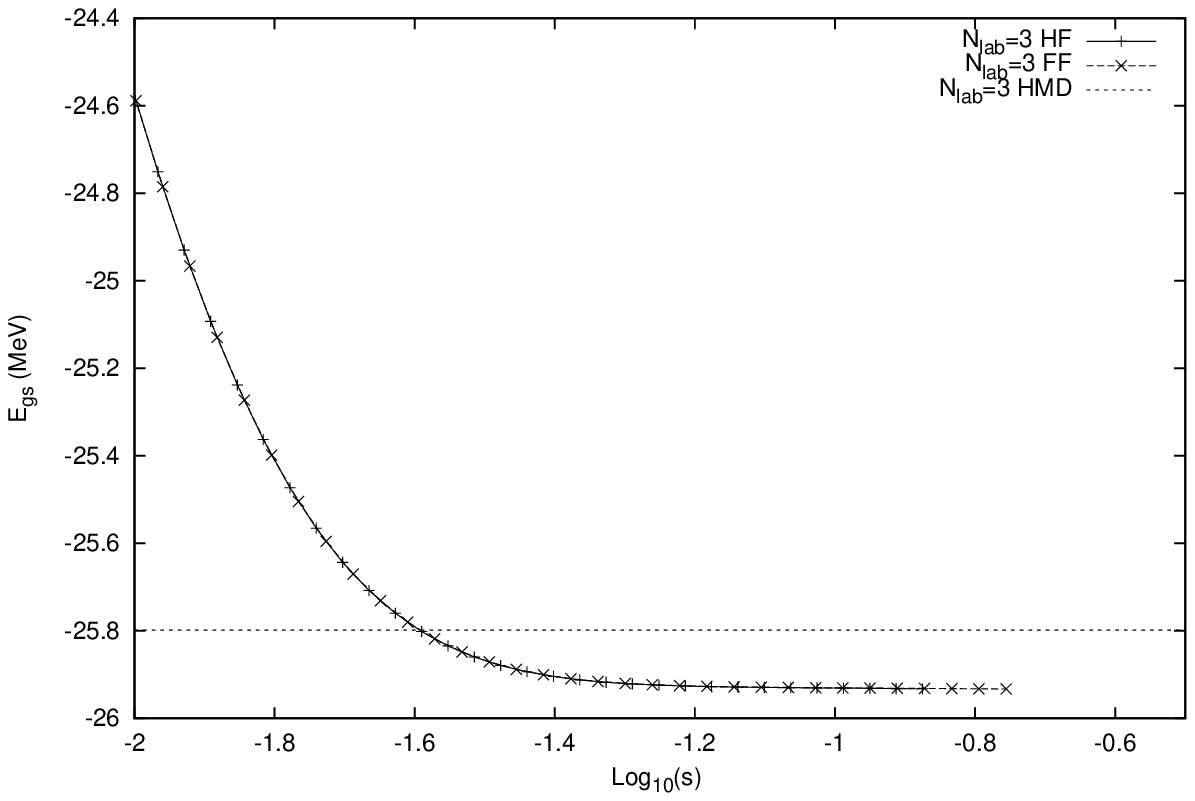}
\caption{${}^4He$ IM-SRG E(s)for large $s$ for the two reference states discussed in the text,
 for $\hbar\om=24 MeV$ and $N_{lab}=3$. The horizontal line is the HMD result.}
\end{figure}
\renewcommand{\baselinestretch}{2}

\renewcommand{\baselinestretch}{1}
\begin{figure}
\centering
\includegraphics[width=10.0cm,height=10.0cm,angle=0]{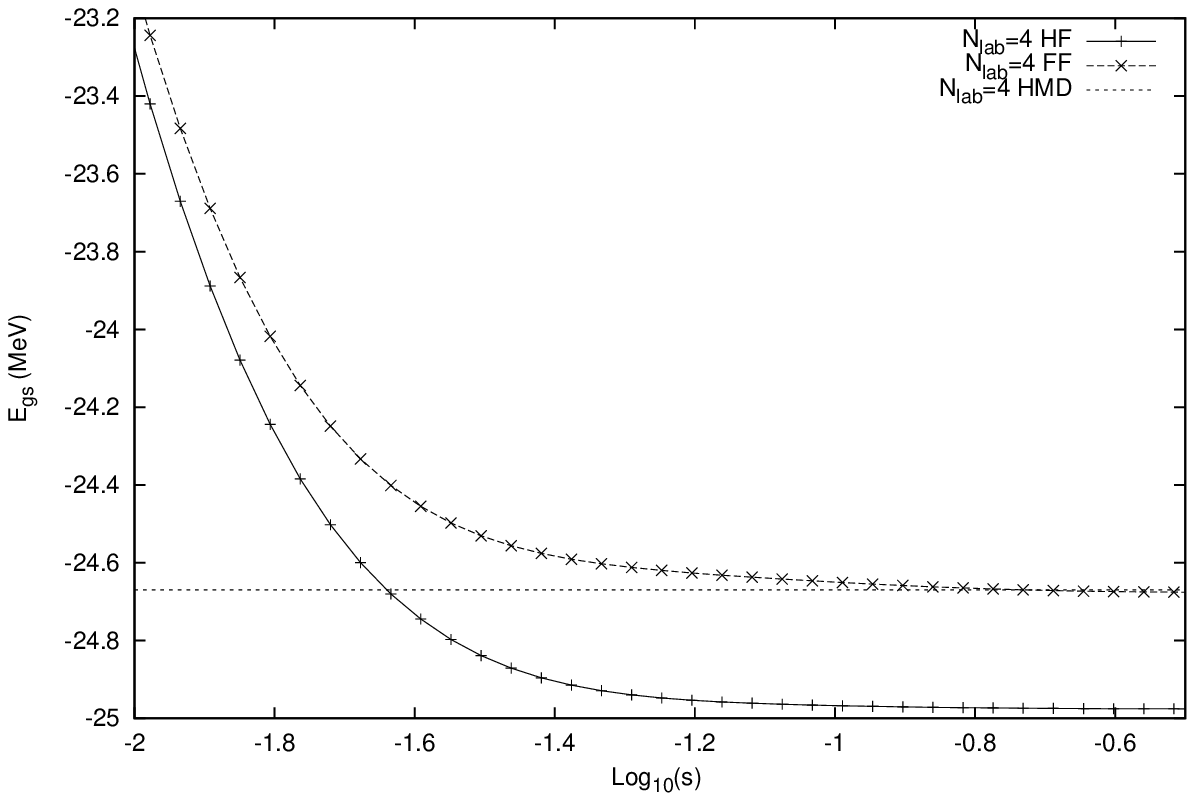}
\caption{${}^4He$ IM-SRG E(s)for large $s$ for the two reference states discussed in the text,
 for $\hbar\om=24 MeV$ and $N_{lab}=4$.  The horizontal line is the HMD result.}
\end{figure}
\renewcommand{\baselinestretch}{2}

\renewcommand{\baselinestretch}{1}
\begin{figure}
\centering
\includegraphics[width=10.0cm,height=10.0cm,angle=0]{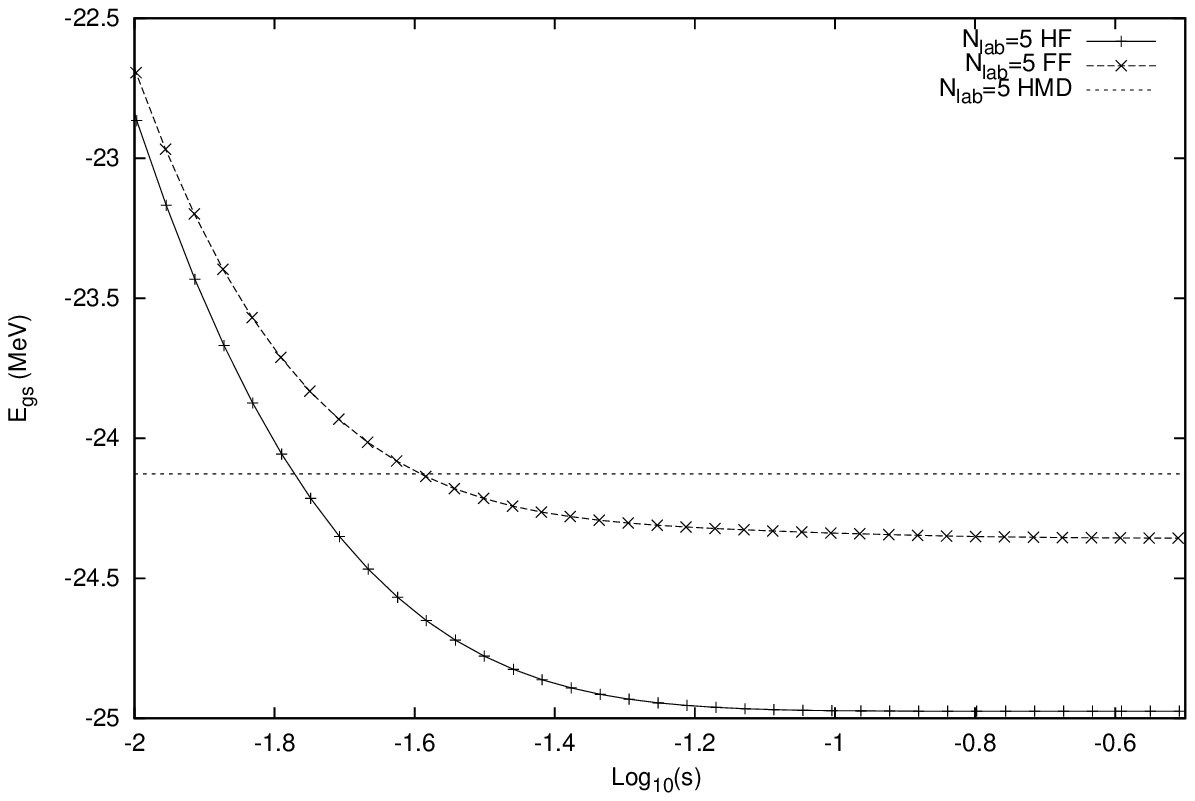}
\caption{${}^4He$ IM-SRG E(s)for large $s$ for the two reference states discussed in the text,
 for $\hbar\om=24 MeV$ and $N_{lab}=5$.  The horizontal line is the HMD result.}
\end{figure}
\renewcommand{\baselinestretch}{2}
\renewcommand{\baselinestretch}{1}
\begin{figure}
\centering
\includegraphics[width=10.0cm,height=10.0cm,angle=0]{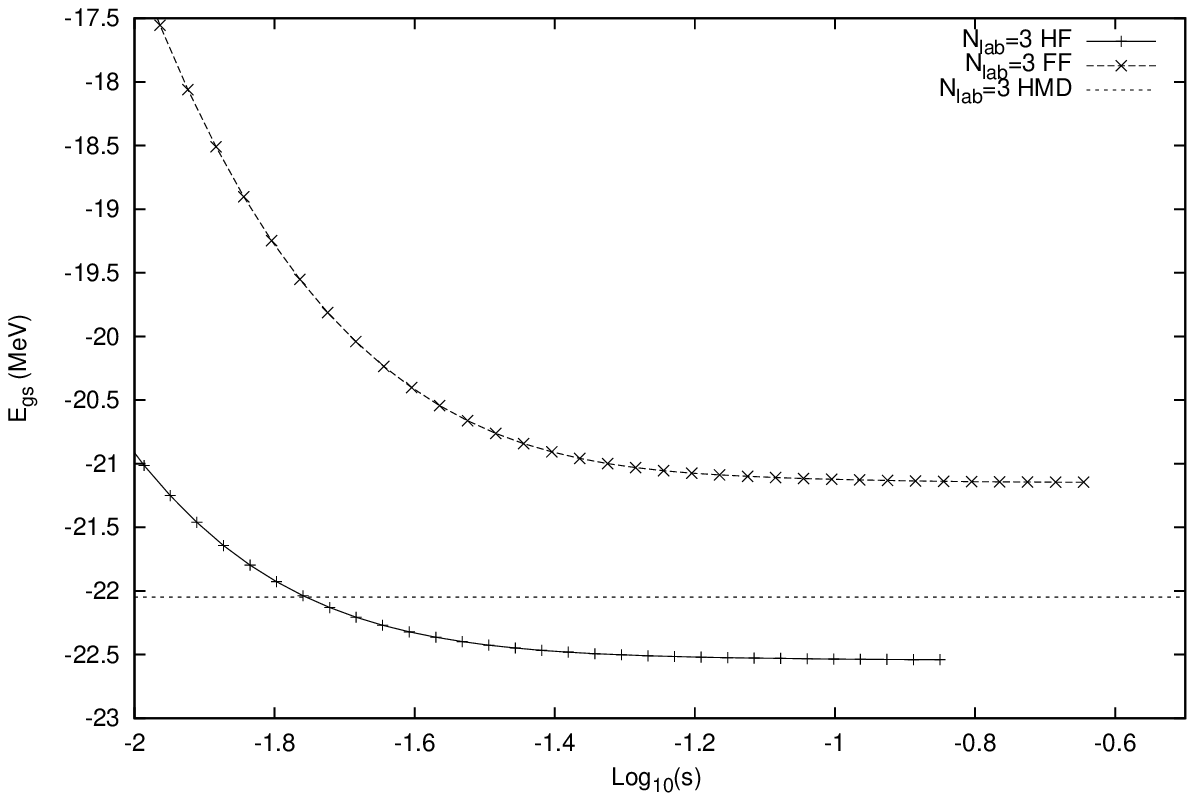}
\caption{${}^4He$ IM-SRG E(s)for large $s$ for the two reference states discussed in the text,
 for $\hbar\om=38 MeV$ and $N_{lab}=3$.  The horizontal line is the HMD result.}
\end{figure}
\renewcommand{\baselinestretch}{2}
\renewcommand{\baselinestretch}{1}
\begin{figure}
\centering
\includegraphics[width=10.0cm,height=10.0cm,angle=0]{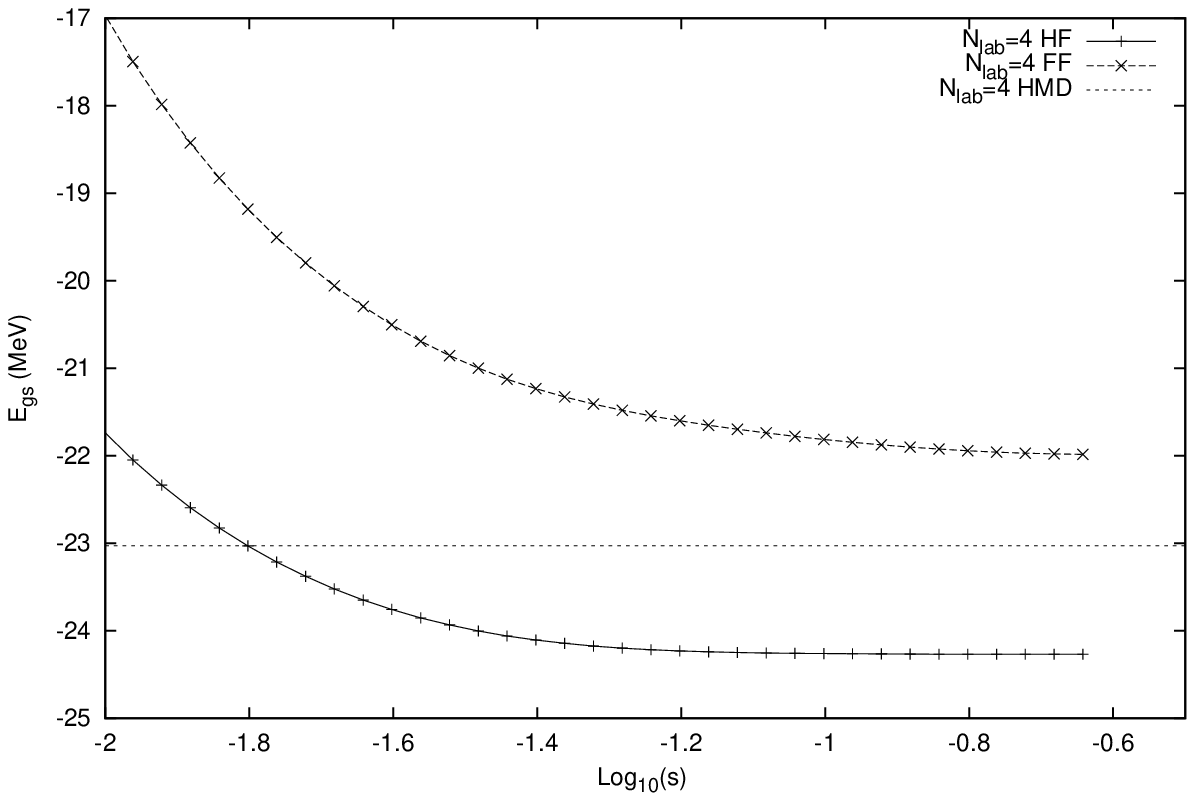}
\caption{${}^4He$ IM-SRG E(s)for large $s$ for the two reference states discussed in the text,
 for $\hbar\om=38 MeV$ and $N_{lab}=4$.  The horizontal line is the HMD result.}
\end{figure}
\renewcommand{\baselinestretch}{2}
\renewcommand{\baselinestretch}{1}
\begin{figure}
\centering
\includegraphics[width=10.0cm,height=10.0cm,angle=0]{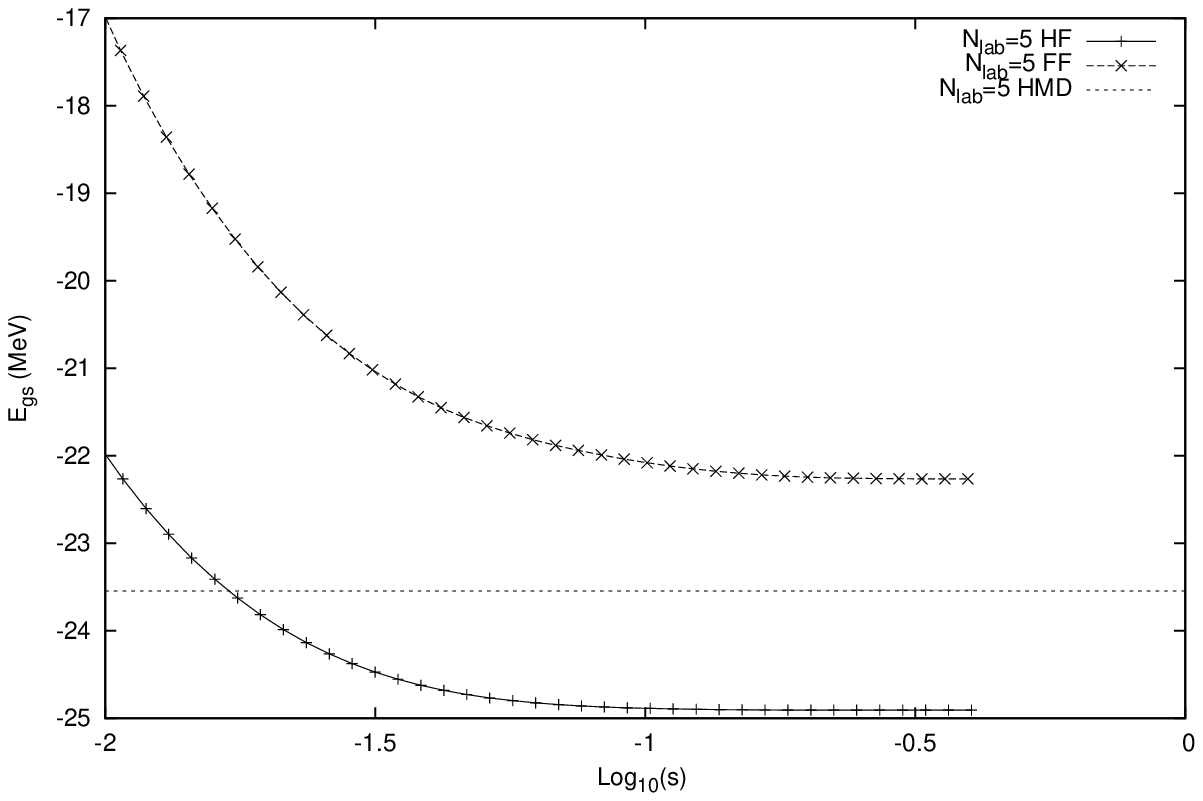}
\caption{${}^4He$ IM-SRG E(s)for large $s$ for the two reference states discussed in the text,
 for $\hbar\om=38 MeV$ and $N_{lab}=5$.  The horizontal line is the HMD result.}
\end{figure}
\renewcommand{\baselinestretch}{2}
\renewcommand{\baselinestretch}{1}
\begin{figure}
\centering
\includegraphics[width=10.0cm,height=10.0cm,angle=0]{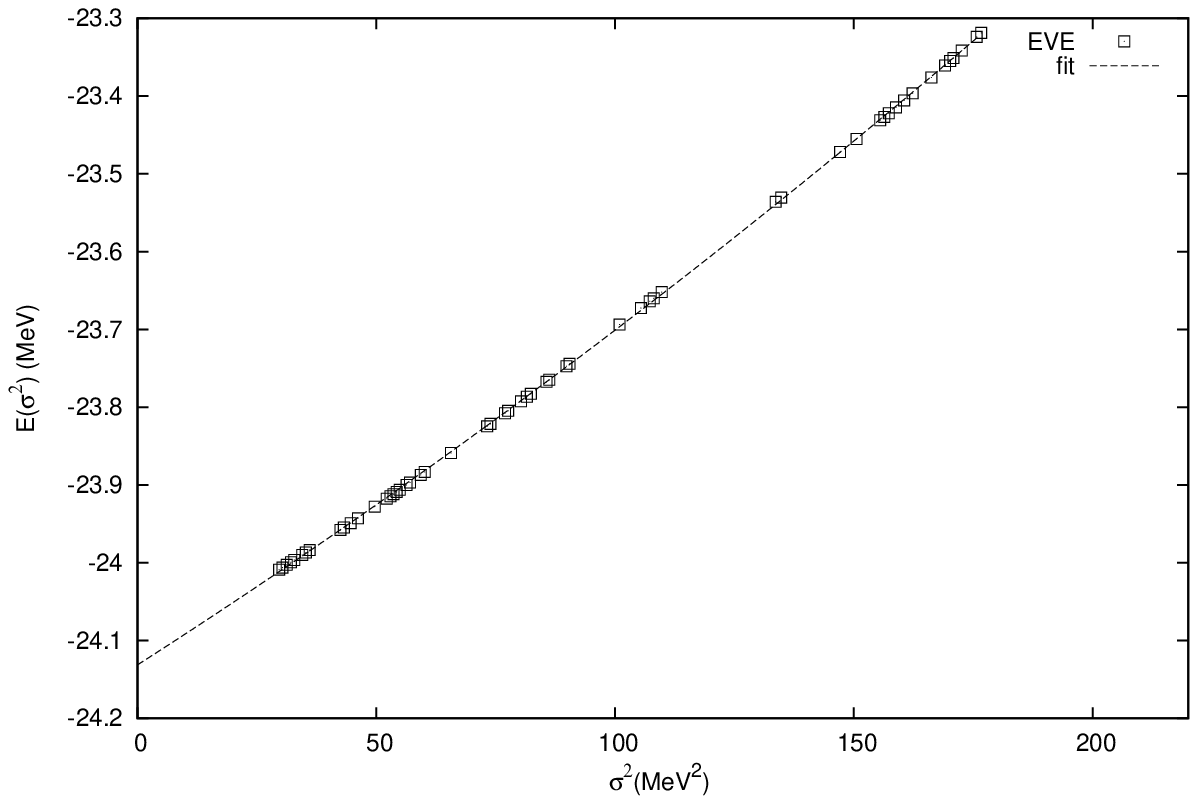}
\caption{EVE plot for ${}^4He$  for $\hbar\om=24 MeV$ and $N_{lab}=5$.}
\end{figure}
\renewcommand{\baselinestretch}{2}

\renewcommand{\baselinestretch}{1}
\begin{figure}
\centering
\includegraphics[width=10.0cm,height=10.0cm,angle=0]{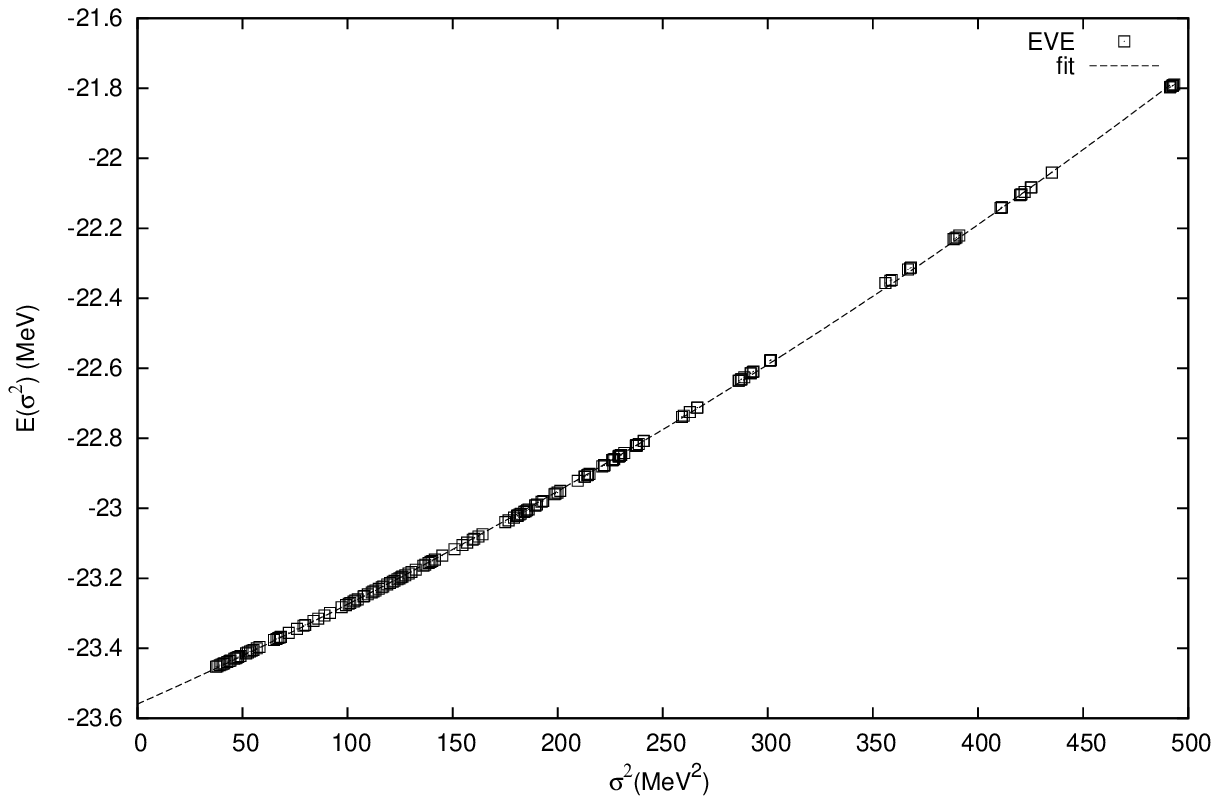}
\caption{EVE plot for ${}^4He$  for $\hbar\om=38 MeV$ and $N_{lab}=5$.}
\end{figure}
\renewcommand{\baselinestretch}{2}
\renewcommand{\baselinestretch}{1}
\begin{figure}
\centering
\includegraphics[width=10.0cm,height=10.0cm,angle=0]{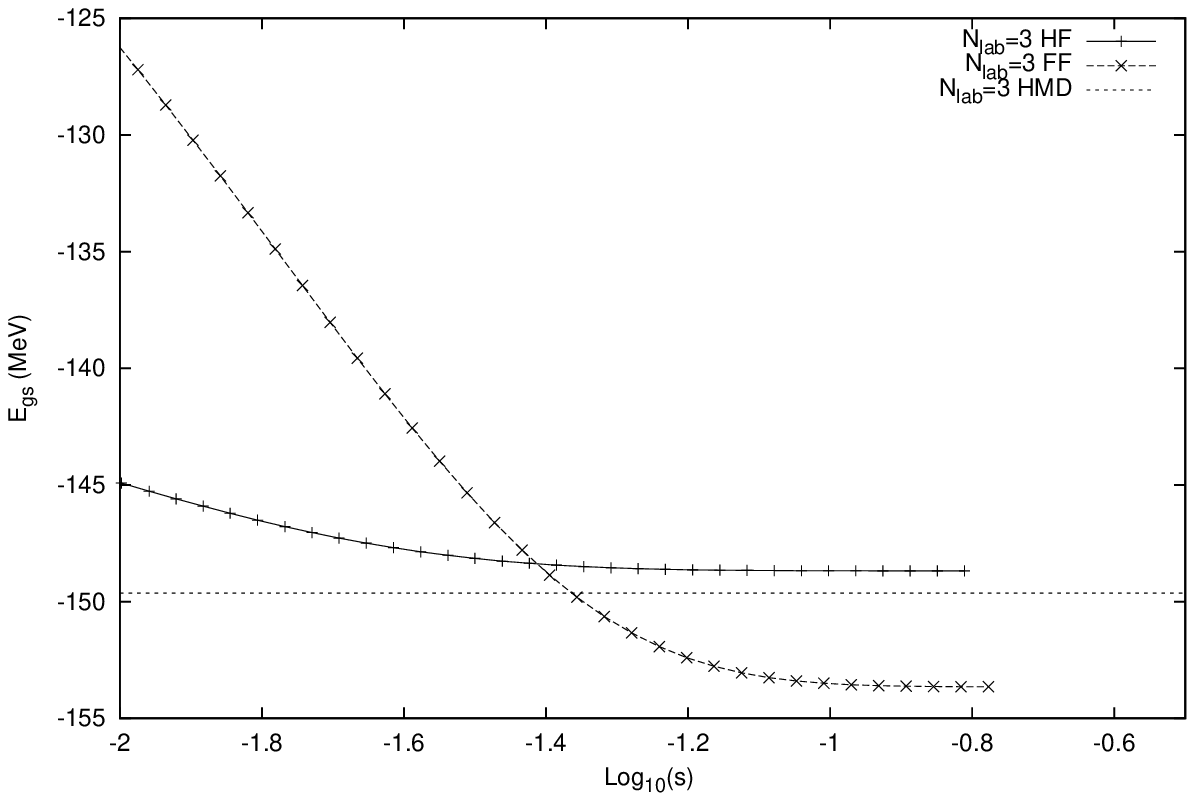}
\caption{IM-SRG results for HF and FF(Fermi filling) for ${}^{16}O$
  at $\hbar\om=14 MeV$ and $N_{lab}=3$ together with the corresponding HMD results.}
\end{figure}
\renewcommand{\baselinestretch}{1}
\begin{figure}
\centering
\includegraphics[width=10.0cm,height=10.0cm,angle=0]{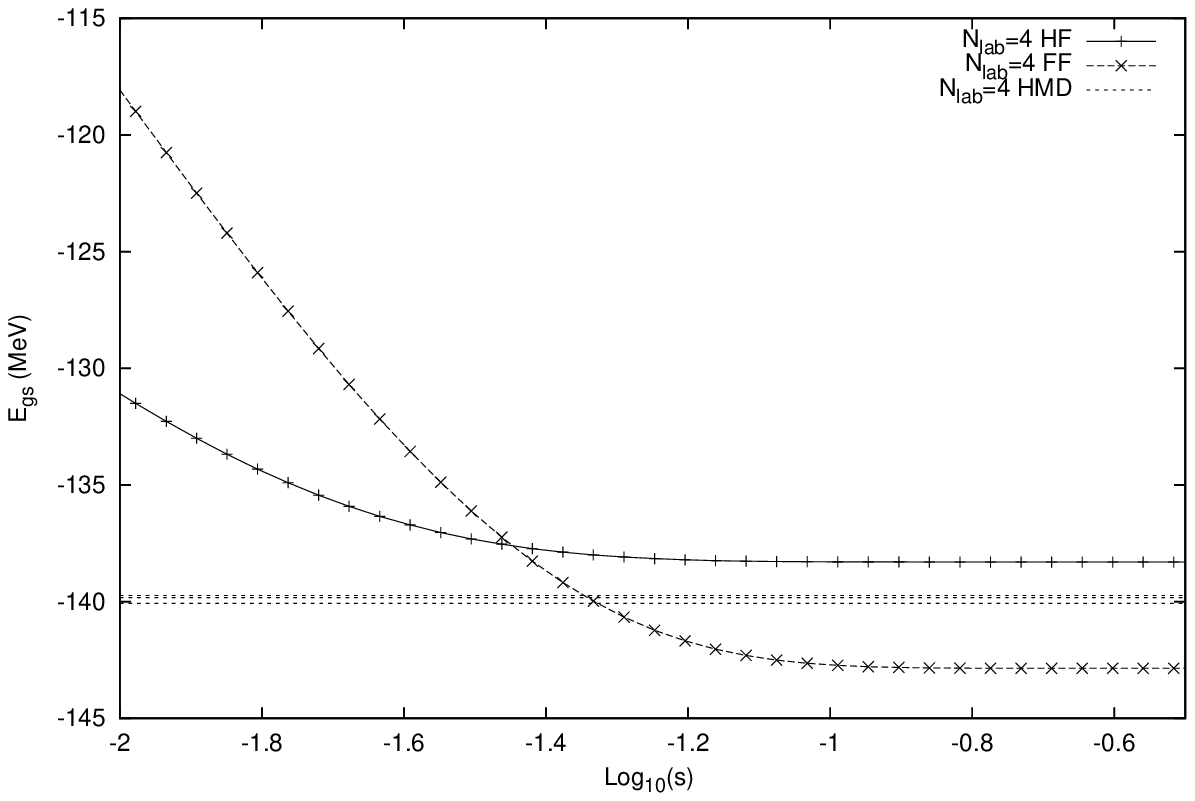}
\caption{IM-SRG results for HF and FF(Fermi filling) for ${}^{16}O$
  at $\hbar\om=14 MeV$ and $N_{lab}=4$ together with the corresponding HMD results.}
\end{figure}
\renewcommand{\baselinestretch}{1}
\begin{figure}
\centering
\includegraphics[width=10.0cm,height=10.0cm,angle=0]{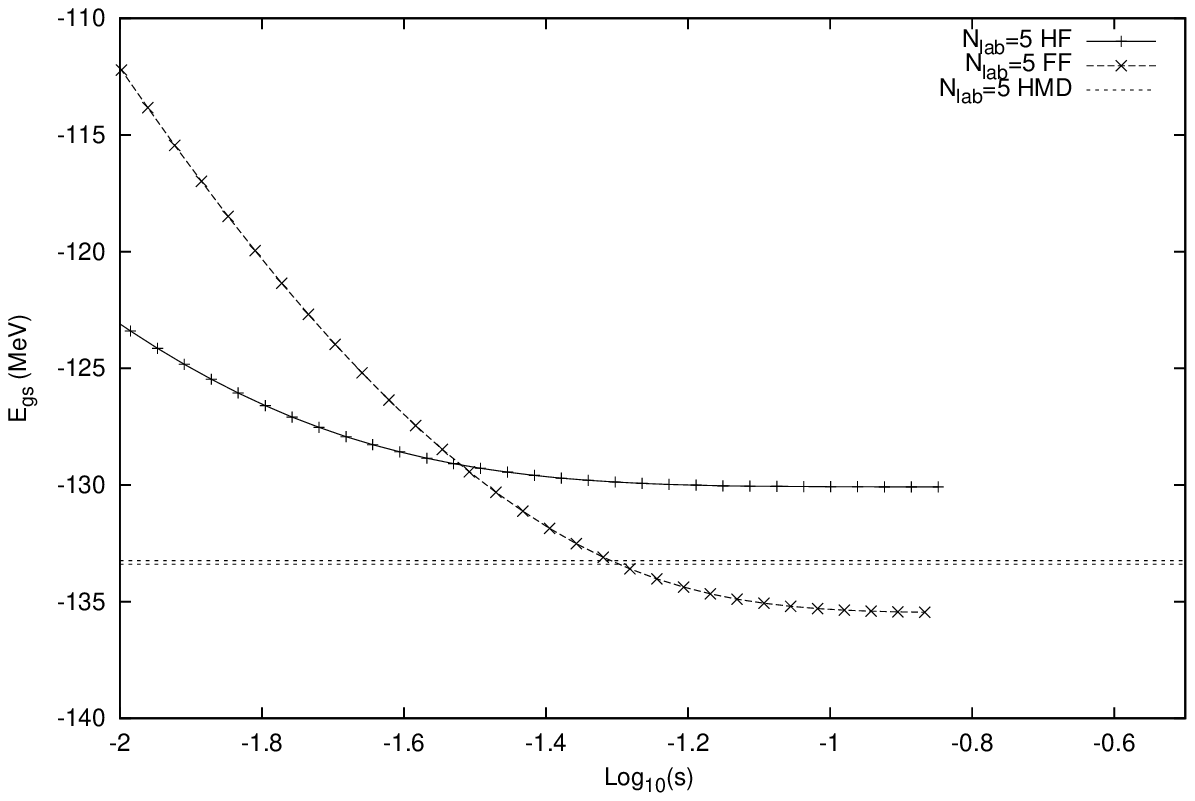}
\caption{IM-SRG results for HF and FF(Fermi filling) for ${}^{16}O$
  at $\hbar\om=14 MeV$ and $N_{lab}=5$ together with the corresponding HMD results.}
\end{figure}
\renewcommand{\baselinestretch}{1}
\begin{figure}
\centering
\includegraphics[width=10.0cm,height=10.0cm,angle=0]{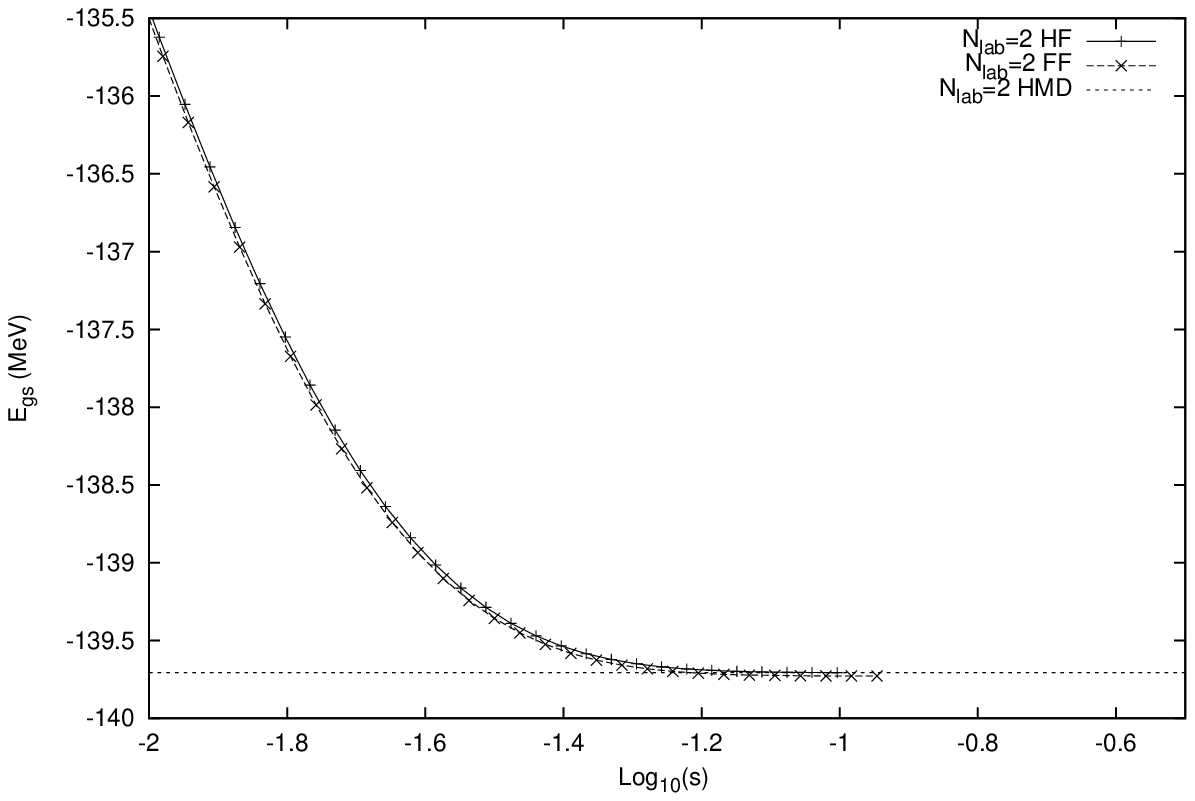}
\caption{IM-SRG results for HF and FF(Fermi filling) for ${}^{16}O$
  at $\hbar\om=24 MeV$ and $N_{lab}=2$ together with the corresponding HMD results.}
\end{figure}

\renewcommand{\baselinestretch}{1}
\begin{figure}
\centering
\includegraphics[width=10.0cm,height=10.0cm,angle=0]{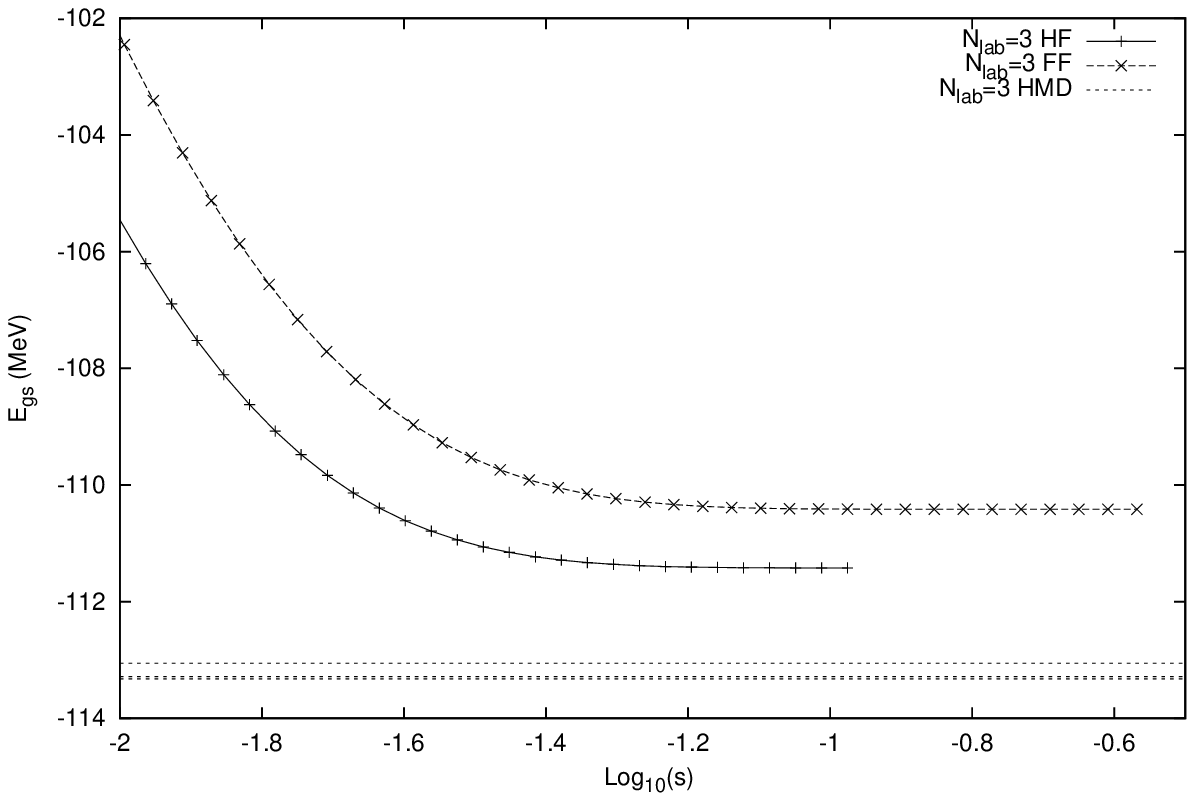}
\caption{IM-SRG results for HF and FF(Fermi filling) for ${}^{16}O$
  at $\hbar\om=24 MeV$ and $N_{lab}=3$ together with the corresponding HMD results.}
\end{figure}
\renewcommand{\baselinestretch}{1}
\begin{figure}
\centering
\includegraphics[width=10.0cm,height=10.0cm,angle=0]{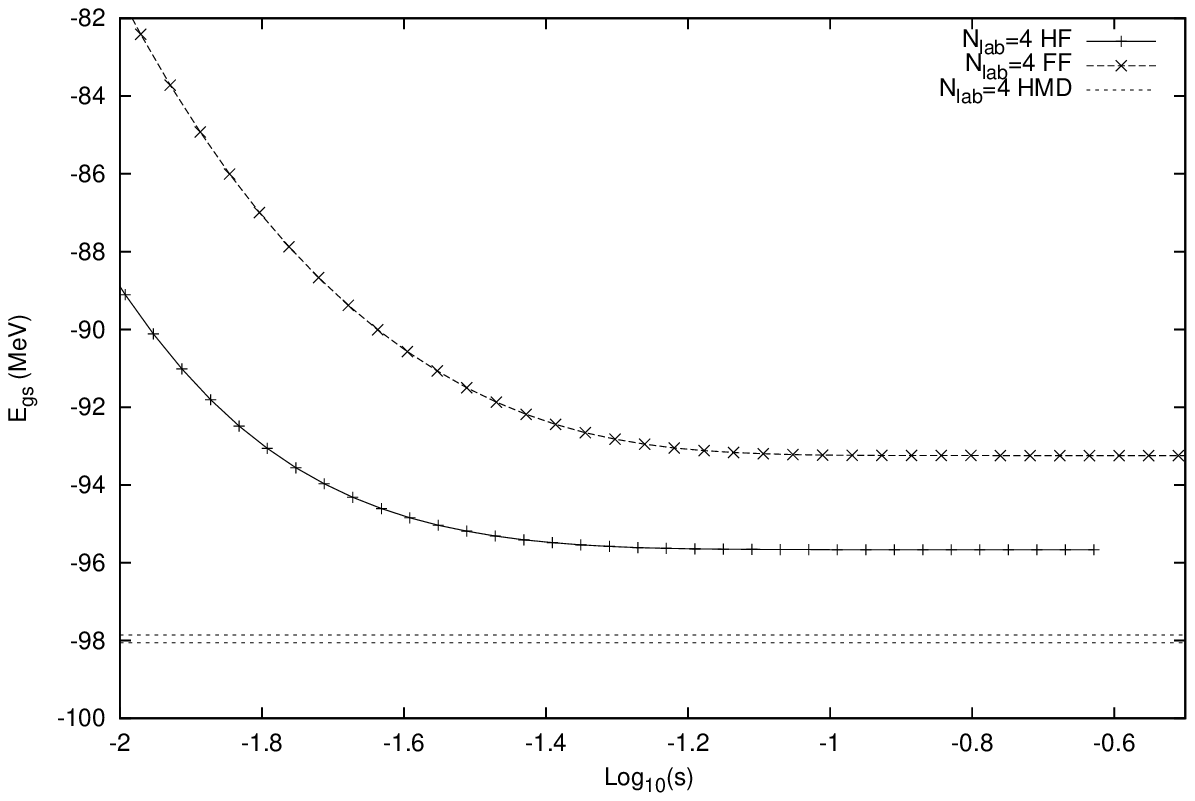}
\caption{IM-SRG results for HF and FF(Fermi filling) for ${}^{16}O$
  at $\hbar\om=24 MeV$ and $N_{lab}=4$ together with the corresponding HMD results.}
\end{figure}

\renewcommand{\baselinestretch}{1}
\begin{figure}
\centering
\includegraphics[width=10.0cm,height=10.0cm,angle=0]{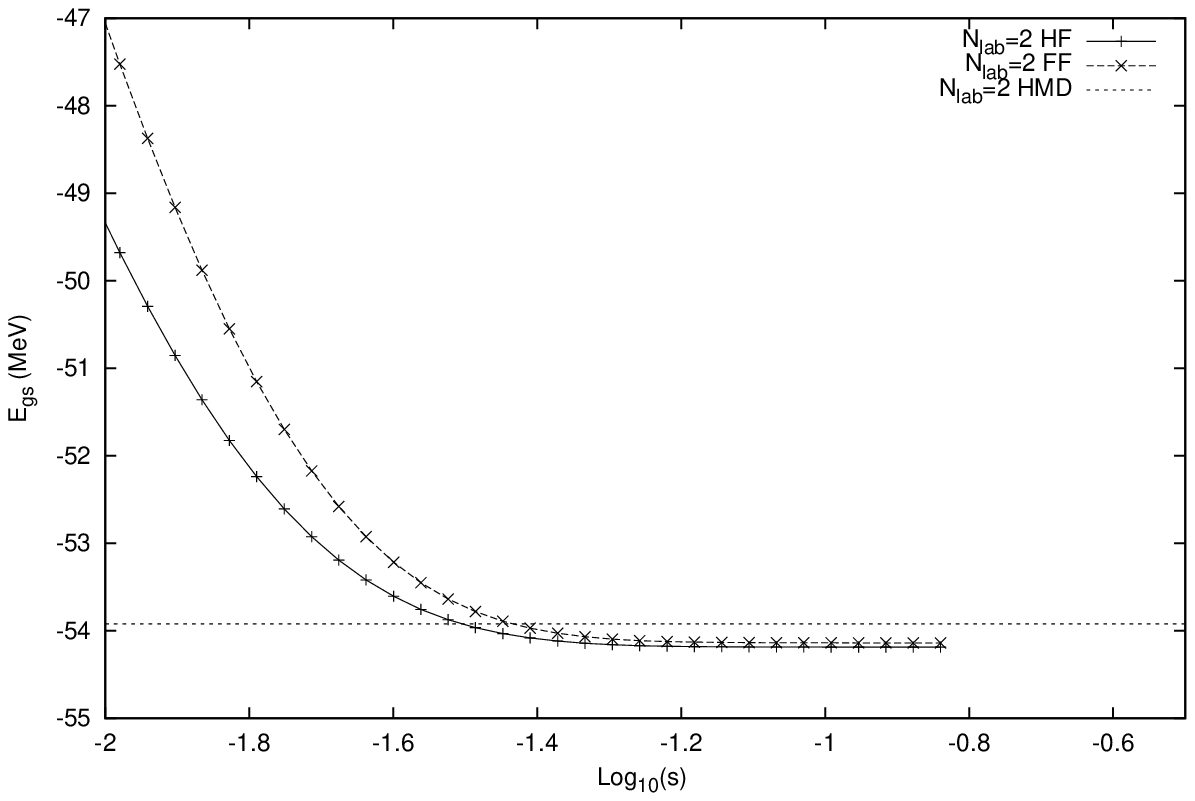}
\caption{IM-SRG results for HF and FF(Fermi filling) for ${}^{16}O$
  at $\hbar\om=32 MeV$ and $N_{lab}=2$ together with the corresponding HMD results.}
\end{figure}

\renewcommand{\baselinestretch}{1}
\begin{figure}
\centering
\includegraphics[width=10.0cm,height=10.0cm,angle=0]{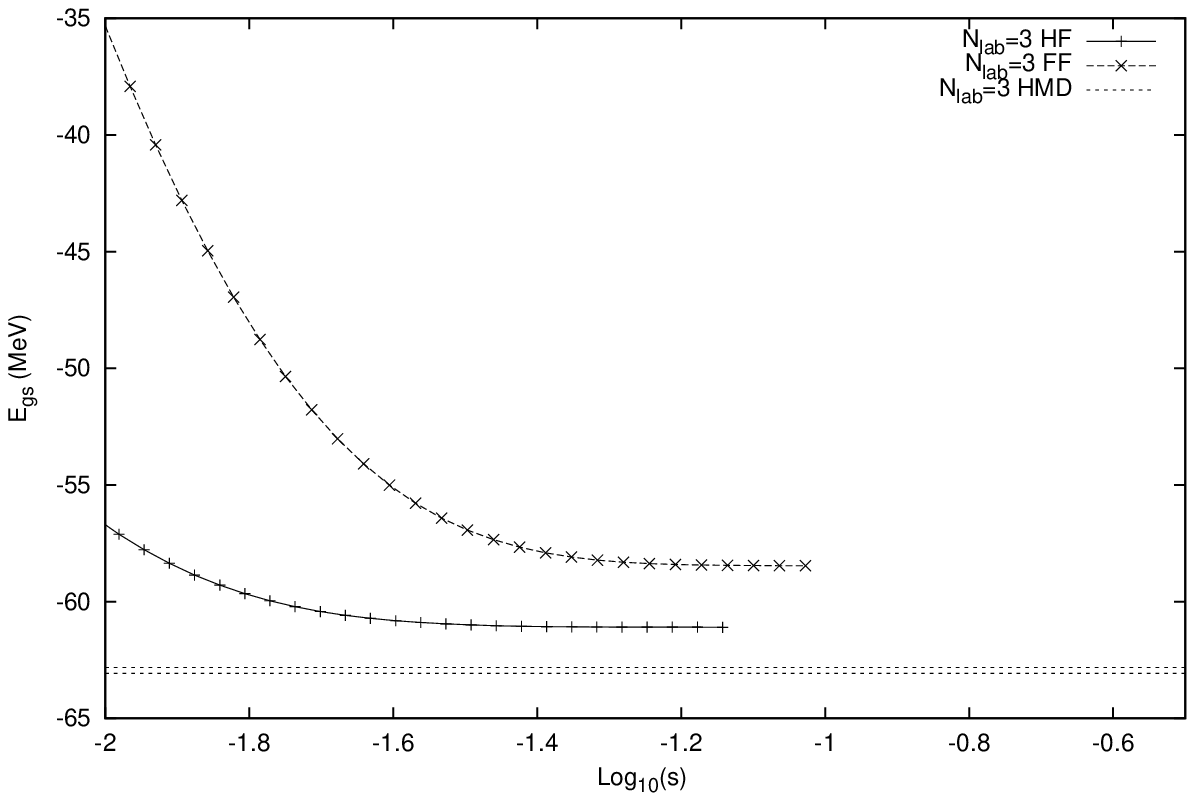}
\caption{IM-SRG results for HF and FF(Fermi filling) for ${}^{16}O$
  at $\hbar\om=32 MeV$ and $N_{lab}=3$ together with the corresponding HMD results.}
\end{figure}
\renewcommand{\baselinestretch}{1}
\begin{figure}
\centering
\includegraphics[width=10.0cm,height=10.0cm,angle=0]{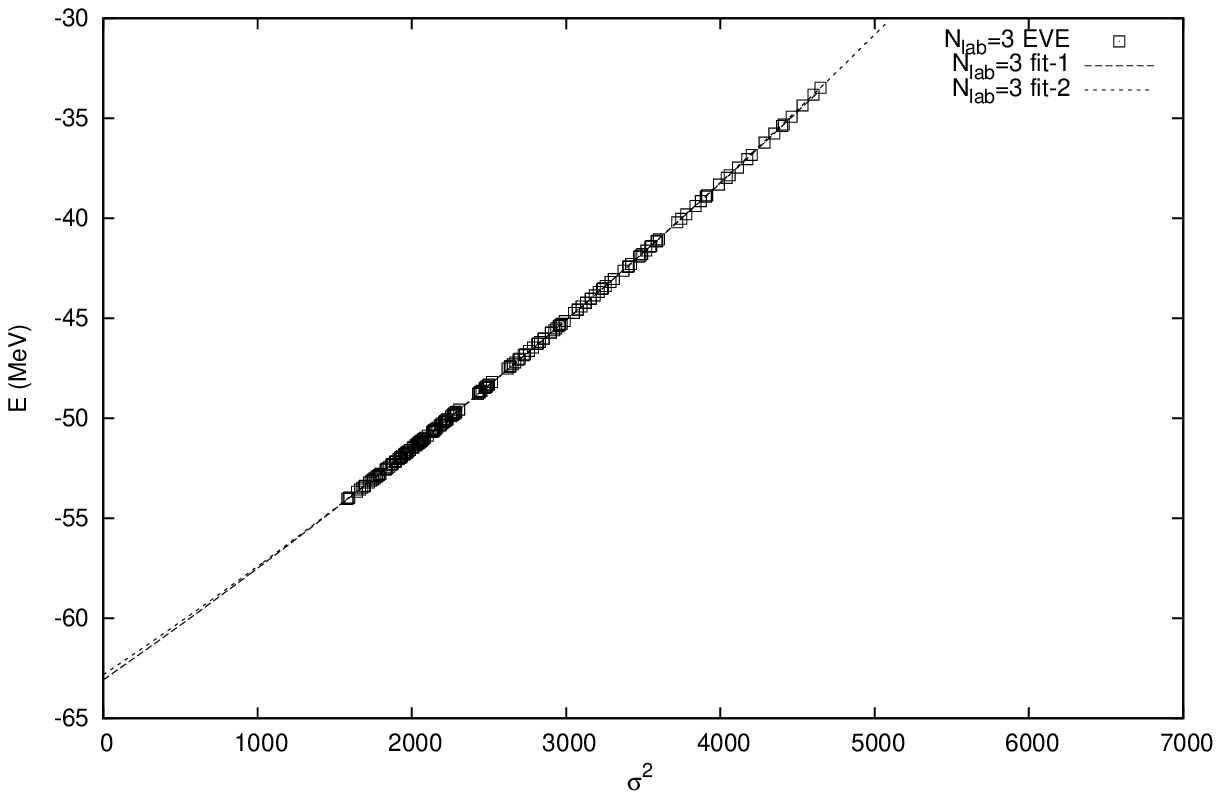}
\caption{EVE results for  ${}^{16}O$,  $\hbar\om=32 MeV$ and $N_{lab}=3$.}
\end{figure}
      In figs.(9)-(11) we plot the end part of the IM-SRG calculations for ${}^{16}O$ at $\hbar\om=14 MeV$
      and $N_{lab}=3,4,5$ respectively together with the corresponding HMD results. The horizontal lines
      represent the HMD results. In some cases we display different fits.
      In figs.(12)-(14) we display the results for $\hbar\om=24 MeV$ and $N_{lab}=2,3,4$,
      and in figs.(15)-(16) the results for $\hbar\om=32 MeV$ and $N_{lab}=2,3$. 
      Notice, as in the case of ${}^4He$, that discrepancies increase with $\hbar\om$ and $N_{lab}$.
      The IM-SRG becomes more and more sensitive to the reference state.
      In order to judge the quality of the EVE extrapolation of the HMD calculations, in fig. (17)
      we show a typical EVE plot $(\sig^2,E(\sig^2)$ and two linear+quadratic fits for ${}^{16}O$
      with $\hbar\om=32 MeV$ and $N_{lab}=3$. Notice that the uncertainty of the fits is determined by 
      the distance of the data from the energy axis. Needless to say by increasing the number
      of Slater determinants we decrease the energy and the variance and, as a consequence, the uncertainty.
      In all plots of the IM-SRG results we decimated the data points in order to avoid clutter.
      In the cases of the EVE extrapolations we select those linear+quadratic fits which have the
      the smallest average discrepancy from the actual calculations.
\par
Before leaving this section, we stress that single-reference
   IM-SRG is usually applied  only to closed shell nuclei.
   For open shell nuclei the more involved multi-reference
   version is preferred. the multi-reference IM-SRG version contains however
   all irreducible rank-2 and rank-3 densities (cf. refs.[9],[38] for more details).
   The importance-truncated NCSM (ref.[39]) can also be used to assess the net effect
   of the many-body forces induced by the IM-SRG flow.
   We limited ourselves to small single-particle spaces, since
   the comparison between single-reference IM-SRG results and exact or
   quasi-exact results is simple. Needless to say, for a comparison
   with experimental data much larger single-particle spaces are needed
   in order to soften the $\hbar\om$ and $N_{lab}$ dependence.
\bigskip
\section {Conclusions.}
\bigskip
      In this work we have performed HMD with EVE extrapolation techniques in order to ascertain
      the overall contribution of the many-body forces induced by  the IM-SRG evolution. They depend on 
      the reference state, on the harmonic oscillator frequency and on the size of the single-particle
      space. Although percentage-wise  these missing many-body forces give a contribution of
      the order of few per cent, their importance seems to increase for increasing harmonic oscillator 
      frequency and with the size of the single-particle space. Most likely they become less relevant
      in the more general multi-reference IM-SRG.
\bigskip
\par
\section {Appendix.}
      We work in the representation that diagonalized the one-body density matrix, 
      i.e. $< \HA^i_j> = n_i\del_{ij}$. We consider only real quantities.
      For completeness we also give the $M-$scheme flow equations (cf.ref.[9] for the more
      general case  of multi-reference state and other generators),
      restricting ourselves to the Brillouin generator given by 
      (we omit for simplicity the $s$ dependence) 
$$
\eta^i_j= f^j_i(n_j-n_i)
\eqno(A1)
$$
$$
\eta^{ij}_{kl}=\Gam^{kl}_{ij}(n_k n_l\bn_i\bn_j-\bn_k \bn_l n_i n_j)
\eqno(A2)
$$
      The gradients of the energy, of the one-body  and of the two-body term are respectively
$$
{d E\over ds }=-\sum_{ab} (\eta^a_b)^2 - {1\over 4}\sum_{abcd} (\eta^{ab}_{cd})^2
\eqno(A3)
$$ 
$$
{d f^i_j\over ds}=\sum_a (\eta^i_a f^a_j-f^i_a\eta^a_j)+
\sum_{ab}(n_a-n_b)(\eta^a_b \Gam^{bi}_{aj}-f^a_b \eta^{bi}_{aj})
$$
$$
+{1\over 2}\sum_{abc} (n_a \bn_b\bn_c+ \bn_a n_b n_c)(\eta^{ia}_{bc}\Gam^{bc}_{ja}-\Gam^{ia}_{bc}\eta^{bc}_{ja})
\eqno(A4)
$$
$$
{d\Gam^{ij}_{kl}\over ds}=\sum_a[(\eta^i_a\Gam^{aj}_{kl}+\eta^j_a\Gam^{ia}_{kl}-
\eta^a_k\Gam^{ij}_{al}-\eta^a_l\Gam^{ij}_{ka})
$$
$$
-(f^i_a\eta^{aj}_{kl}+f^j_a\eta^{ia}_{kl}-
f^a_k\eta^{ij}_{al}-f^a_l\eta^{ij}_{ka})]
$$
$$
+ {1\over 2}\sum_{ab}(1-n_a-n_b)(\eta^{ij}_{ab}\Gam^{ab}_{kl}- \Gam^{ij}_{ab}\eta^{ab}_{kl})
$$
$$
+\sum_{ab}(n_a-n_b)[( \eta^{ia}_{kb}\Gam^{jb}_{la}-\Gam^{ia}_{kb}\eta^{jb}_{la})
-                   ( \eta^{ja}_{kb}\Gam^{ib}_{la}-\Gam^{ja}_{kb}\eta^{ib}_{la})]
\eqno(A5)
$$
      In the above equations all single-particle indices comprise the $m$ quantum numbers and
      $\bn=1-n$.
      The last term in eq.(A5) is referred to as the cross-term, and it is the most troublesome
      in going to the angular momentum coupled representation.
      The variational character of the Brillouin generator is apparent in this
      form of the energy gradient.
\par
      In the coupled representation (the $J-$flow) and for closed shells
      the single-particle labels comprise the radial, the
      angular momentum and the isospin quantum numbers only, e.g. $i=(k_i,l_i,j_i,\tau_i)$.
      All one-body
      quantities are diagonal in the $l,j,\tau$ quantum numbers. 
      The Brillouin generators become
$$
\eta^i_j= f^j_i(n_j-n_i)
\eqno(A6)
$$
$$
\eta^{ij}_{kl}(J)=\Gam^{kl}_{ij}(J)(n_k n_l\bn_i\bn_j-\bn_k \bn_l n_i n_j)
\eqno(A7)
$$
      for the one-body and two-body parts respectively.
      $J$ is the total two-body angular momentum in which two-body matrix elements are
      diagonal and $M$ independent. We use the notation $\hj=\sqrt{2j+1}$.
      The energy gradient ${d E\over ds}$ is given by
$$
{dE\over ds}=-\sum_{ab}\de_{j_a,j_b}\de_{l_a,l_b}\de_{\tau_a,\tau_b} \hj_a^2 (\eta^a_b)^2-
 {1\over 4} \sum_{abcd J} \HJ^2 [\eta^{ab}_{cd}(J)]^2
\eqno(A8)
$$
      The $\de$'s are Kronecker deltas.
      The gradient of the one-body term is given by
$$
{d f^i_j\over ds}=  \sum_a \de_{j_i,j_a} (\eta^i_a f^a_j-f^i_a \eta^a_j)
+{1\over \hj_i^2} \sum_{abJ}\HJ^2(n_a-n_b)[\eta^a_b\Gam^{bi}_{aj}(J)-f^a_b\eta^{bi}_{aj}(J)]
$$
$$
+{1\over 2\hj_i^2}\sum_{abcJ}\HJ^2(n_a\bn_b\bn_c+\bn_a n_b n_c)[\eta^{ia}_{bc}(J)\Gam^{bc}_{ja}(J)-
\Gam^{ia}_{bc}(J)\eta^{bc}_{ja}(J)]
\eqno(A9)
$$
     The flow equation for $\Gam $ is more involved. For readability, we separate contributions to 
     the gradient 
     of $\Gam$ in three terms 
$$
{d\Gam^{ij}_{kl}(J)\over ds} = C^{ij}_{kl}(J)+D^{ij}_{kl}(J)+E^{ij}_{kl}(J)
\eqno(A10)
$$
     where
$$ 
C^{ij}_{kl}(J)= \sum_a \{[ \eta^i_a\Gam^{aj}_{kl}(J)-f^i_a\eta^{aj}_{kl}(J)]+
                            [\eta^j_a\Gam^{ia}_{kl}(J)-f^j_a\eta^{ia}_{kl}(J)]-
$$
$$
                      [\eta^a_k\Gam^{ij}_{al}(J)-f^a_k\eta^{ij}_{al}(J)]-
                      [\eta^a_l\Gam^{ij}_{ka}(J)-f^a_l\eta^{ij}_{ka}(J)] \}
\eqno(A11)
$$
     and
$$
D^{ij}_{kl}(J)={1\over 2}\sum_{ab}(1-n_a-n_b)[\eta^{ij}_{ab}(J)\Gam^{ab}_{ij}(J)-
                          \Gam^{ij}_{ab}(J)\eta^{ab}_{ij}(J)]
\eqno(A12)
$$
    The cross-coupled term $E$ is more involved.
    Define first
$$
W^{ij}_{kl}(J)= (-1)^{J+j_j+j_l}\sum_{ab}(n_a-n_b)\sum_K \HK^2
\bigg\{\matrix{ j_i& j_j & J \cr j_l & j_k  & K }\bigg\}\times
$$
$$
[\tilde \eta_K (k\overline i,b\overline a)\tilde\Gam_K(l\overline j,a\overline b)-
\tilde  \Gam_K(k\overline i,b\overline a)\tilde\eta_K(l\overline j,a\overline b) ]
\eqno(A13)
$$ 
   where the tilde matrices are defined as
$$
\tilde O_K(l\overline j,a \overline b) = \sum_J (-1)^J \HJ^2 O^{jb}_{la}(J)
 \bigg\{\matrix{ j_j & j_l  & K\cr j_a & j_b & J}\bigg\} 
\eqno(A14)
$$
     The last contribution to $ d\Gam^{ij}_{kl}(J)/ds$ is then given by
$$ 
E^{ij}_{kl}(J) =W^{ij}_{kl}(J)-(-1)^{j_i+j_j+J} W^{ji}_{kl}(J)
\eqno(A15)
$$
    Apart a phase factor, the tilde operators are the Pandya transform of the corresponding ones
    without the tilde.

\vfill
\eject
\end{document}